\documentclass[preprint,aps]{revtex4}

\usepackage{graphicx}

\begin{document}

\title{Phase Diagram and High Temperature Superconductivity at 65 K in Tuning Carrier Concentration of Single-Layer FeSe Films}
%%\title{Competing Phases and Optimization of High-Temperature Superconductivity in Tuning Carrier Concentration of the Single-Layer FeSe Thin Film}
%%\title{Competing Phases and Optimization of High-Temperature Superconductivity in Tuning Charge Carriers of the Single-Layer FeSe Thin Films}
%%\title{Competing Phases and Emergence of High-Temperature Superconductivity in Tuning Charge Carriers of the Single-Layer FeSe Thin Films}

\author{Shaolong He$^{1,\sharp}$, Junfeng He$^{1,\sharp}$, Wenhao Zhang$^{2,3,\sharp}$, Lin Zhao$^{1,\sharp}$, Defa Liu$^{1}$,  Xu Liu$^{1}$, Daixiang Mou$^{1}$,   Yun-Bo Ou$^{3}$, Qing-Yan Wang$^{2,3}$, Zhi Li$^{3}$, Lili Wang$^{3}$,  Yingying Peng$^{1}$, Yan Liu$^{1}$, Chaoyu Chen$^{1}$, Li Yu$^{1}$, Guodong  Liu$^{1}$,  Xiaoli Dong$^{1}$, Jun Zhang$^{1}$, Chuangtian Chen$^{4}$, Zuyan Xu$^{4}$,  Xi Chen$^{2}$, Xucun Ma$^{3,*}$,  Qikun Xue$^{2,*}$,   and X. J. Zhou$^{1,*}$ }

\affiliation{
\\$^{1}$National Lab for Superconductivity, Beijing National Laboratory for Condensed Matter Physics, Institute of Physics,
Chinese Academy of Sciences, Beijing 100190, China
\\$^{2}$State Key Lab of Low-Dimensional Quantum Physics, Department of Physics, Tsinghua University, Beijing
100084, China
\\$^{3}$Beijing National Laboratory for Condensed Matter Physics, Institute of Physics,
Chinese Academy of Sciences, Beijing 100190, China
\\$^{4}$Technical Institute of Physics and Chemistry, Chinese Academy of Sciences, Beijing 100190, China
}
\date{July 30, 2012}
%
%The abstract goes here

\begin{abstract}

Superconductivity in the cuprate superconductors and the Fe-based superconductors is realized by doping the parent compound with charge carriers, or by application of high pressure, to suppress the antiferromagnetic state. Such a rich phase diagram is important in understanding superconductivity mechanism and other physics in the Cu- and Fe-based high temperature superconductors.  In this paper, we report a phase diagram  in the single-layer FeSe films grown on SrTiO$_3$ substrate by an annealing procedure to tune the charge carrier concentration over a wide range. A dramatic change of the band structure and Fermi surface is observed, with two distinct phases identified that are competing during the annealing process.  Superconductivity with a record high transition temperature (T$_c$) at 65$\pm$5 K is realized by optimizing the annealing process. The wide tunability of the system across different phases, and its high-T$_c$, make the single-layer FeSe film ideal not only to investigate the superconductivity physics and mechanism, but also to study novel quantum phenomena and for potential applications.

\end{abstract}

%%%%\pacs{74.70.-b, 74.25.Jb, 79.60.-i, 71.20.-b}

\maketitle

In  high temperature cuprate superconductors, superconductivity is realized by doping the parent Mott insulator with charge carriers to suppress the antiferromagnetic state\cite{CuHTSC}. In the process, the physical property experiences a dramatic change from antiferromagnetic insulator, to a superconductor and eventually to a non-superconducting normal metal. In the superconducting region, the transition temperature T$_c$ can be tuned by the carrier concentration, initially going up with the increasing doping, reaching a maximum at an optimal doping, and then going down with further doping\cite{CuHTSC}.  Such a rich evolution with doping not only provides a handle to tune the physical properties in a dramatic way, but also provides clues and constraints in understanding the origin of the high-T$_c$ superconductivity.  The same is true for the Fe-based superconductors where superconductivity is achieved by doping the parent magnetic compounds which are nevertheless metallic\cite{FeHTSC1,FeHTSC2}. Again, the superconducting transition temperature can be tuned over a wide doping range with an maximum T$_c$ at the optimal doping. Understanding such a rich evolution is also a pre-requisite in understanding the origin of high temperature superconductivity in the Fe-based superconductors.

The latest discovery of high temperature superconductivity signature in the single-layer FeSe films\cite{QKXue,DFLiu} is significant in a couple of respects. First, it may exhibit a high T$_c$ that breaks the T$_c$ record ($\sim$55 K) in the Fe-based superconductors kept so far since 2008\cite{Kamihara,ZARenSm,RotterSC,MKWu11,CQJin111,JGGuo}. Second, the discovery of such a high-T$_c$ in the single-layer FeSe film is surprising when considering that its bulk counterpart has a T$_c$ only at 8 K\cite{MKWu11} although it can be enhanced to 36.7 K under high pressure\cite{FeSeHighP}. Third, it provides an ideal system to investigate the origin of high temperature superconductivity. On the one hand, this system consists of a single-layer FeSe film that has a simple crystal structure and strictly two-dimensionality; its simple electronic structure may provide key insights on the high T$_c$ superconductivity mechanism in the Fe-based compounds\cite{DFLiu}. On the other hand, the unique properties of this system may involve the interface between the single-layer FeSe film and the SrTiO$_3$  substrate that provides an opportunity to investigate the role of interface in generating high-T$_c$ superconductivity\cite{QKXue}.

Like in cuprates and other Fe-based superconductors, it is important to explore whether one can tune the single-layer FeSe system to vary its physical properties and superconductivity by changing the charge carrier concentration.  In this paper, we report a wide range tunability of the electronic structure and physical properties that is realized in the single-layer FeSe film. Instead of chemical substitution that is commonly used in cuprates and most of other Fe-based superconductors, the doping here is realized by a simple annealing process. The electronic structure variation with the annealing process is dramatic, varying from one end phase (called N phase) to the other (called S phase), with their competition in the intermediate stage. Signature of superconductivity is observed in the S phase, which varies with doping and when optimized, can reach a record high T$_c$ at (65$\pm$5) K. Such a wide tunability of the single-layer FeSe film not only  provides an opportunity to investigate the superconductivity physics and mechanism of the Fe-based superconductors, it also provides an ideal system to fabricate heterostructure devices for novel quantum phenomena and potential applications.

The as-prepared single-layer FeSe film grown at relatively low temperature is non-superconducting; superconductivity is realized by the subsequent annealing in vacuum at a relatively high temperature for a period of time\cite{QKXue,DFLiu}.  To keep track on the evolution from the non-superconducting to superconducting transition during the annealing process,  we have divided the annealing process into many small steps by gradually increasing the annealing temperature and annealing time (see Supplementary Fig. S1). The samples were {\it in situ} annealed, and then measured by angle-resolved photoemission (ARPES) on the band structure, Fermi surface and energy gap after each annealing step (see Supplementary Materials and  Methods).  We note that due to slight variation in the SrTiO$_3$ substrates and the initial preparation conditions of the single-layer FeSe thin films, the starting point of the electronic structure may vary between samples. However, their annealing process all follow the typical trend that are shown in Figs. 1 and 2 and eventually come to the similar ending point.  We have worked on many samples and the results are highly reproducible.

Figure 1 shows the Fermi surface evolution with the annealing process for two single-layer FeSe samples at different annealing stages; they basically span the entire annealing process we can have under our annealing conditions. The corresponding band structures are shown in Fig. 2. For convenience, we use the annealing sequence to denote different samples annealed under different conditions, as indicated in the Supplementary Fig. S1. As seen in Figs. 1 and 2, the Fermi surface and the band structure of the single-layer FeSe film experience a dramatic change during the annealing process.  One may wonder whether there is any signal that may come from the SrTiO$_3$ substrate. To check on this point, we have prepared a SrTiO$_3$ substrate using the same procedure as that in making single-layer FeSe thin films, and then annealed the substrate by the similar process as used for the FeSe films.  Our careful ARPES measurements indicate that none of the bands reported in the paper come from the SrTiO$_3$ substrate prepared in such a process.

There are two distinct phases appeared during the annealing process, as seen from Figs 1 and 2. In the initial stage, the electronic structure of the first three sequences (1 to 3) is similar which can be attributed to a pure N phase, with a schematic band structure shown in Fig. 2d (the detailed band structure of the N phase and its corresponding photoemission spectra are shown in the Supplementary Fig. S2). The spectral distribution as a function of momentum is mainly characterized by two ``strong spots'' near M point. There is also some spectral weight near the zone center $\Gamma$ due to the N1 band (Fig. 2d) because its top is quite close to the Fermi level. Further annealing leads to the appearance of additional bands, which start from the sequence 4 and get more and more pronounced with annealing. In the mean time, the bands corresponding to the N phase decrease in intensity and completely become invisible till the sequence 10 (Fig. 2(a-c)). The band structure for the sequence 10 and thereafter are similar and can be ascribed to another pure S phase which has a schematic band structure shown in Fig. 2e (the detailed band structure of the S phase and its corresponding photoemission spectra are shown in Supplementary Fig. S3). The S phase is characterized by an electron-like Fermi surface around the M point (Fig. 1)\cite{DFLiu}.  It appears that the N phase is stable at relatively low annealing temperature  while the S phase becomes dominant at relatively high annealing temperature. In the intermediate stage, the band structure can be understood as a mixture of the N phase and S phase (the detailed electronic structure of a mixed phase can be seen from the Supplementary Fig. S4). The corresponding Fermi surface is also a combination of the N and S phases, with both ``strong spots" and electron-like Fermi surface near the M point (most notable for the sequence 4 and 5 in Fig. 1). Although these two phases can coexist in the intermediate stage, they compete with each other in the sense that the S phase increases in intensity with annealing at the expense of the N phase.

In addition to the evolution between the N phase and the S phase, the band structure for a given phase also changes during the annealing process (Fig. 2(a-c)), signaling a change of the carrier concentration in these two phases.  Furthermore, the band structure change shows different trends and magnitudes with the annealing process as plotted in Figs. 2f and 2g. These indicate that the doping process in both the N phase and the S phase is not a rigid-band shift. For the S phase, the doping concentration can be estimated based on the Fermi surface size around  M point. Assuming two degenerate electron-like Fermi surface sheets near M, the estimated doping of the S phase is shown in Fig. 2h. The S phase is electron-doped and its concentration gradually increases with the annealing process. On the other hand, the effective mass for the electron-like band near M,  which can be determined from the band width and the Fermi momentum, is around 3m$_e$ (where m$_e$ represents a static mass of an electron) and shows little change with doping.

One immediate question arises is what is really happening during the annealing process. The change-over from one end phase to the other, and the concomitant change of the carrier concentration in a given phase, all indicate that the annealing process has effectively caused a dramatic change on the FeSe sample. Since the annealing process is rather mild that occurs in vacuum at relatively low temperature, there are a couple of possible scenarios. The first is the loss of Se because Se is easy to evaporate; in this case, its loss would give rise to electron doping in the FeSe$_{1-x}$ system. Indication of Se vacancy in the FeSe film grown on graphene substrate is observed before from the scanning electron microscope (STM) measurement (Fig. 3c in \cite{CLSongPRB}). Future work can be done to check whether one can identify such Se vacancy in the FeSe film grown on SrTiO$_3$ substrate and its variation during the annealing process.  The second possibility is the loss of oxygen near the SrTiO$_3$ surface; it will also lead to electron doping which can be transferred to FeSe near the interface. Whether the oxygen loss can be significant enough at such a low annealing temperature remains to be investigated. The re-arrangement of Fe and Se atoms on the SrTiO$_3$ surface during the annealing can also be a possibility. The exact working mechanism of the annealing process needs further experimental and theoretical efforts to sort out.

The band structure and the spectral weight distribution of the N phase near the M point exhibit clear resemblance to that of the parent AEFe$_2$As$_2$ (AE=Ba or Sr) compounds in the magnetic state\cite{Feng122,HYLiuSr,GDLiuBa,PRichard} (see Supplementary Fig. S5). (1). The band structure of the N phase is characterized by a hole-like band near M, labeled as N2 in Fig. 2d. Similar hole-like band is also observed in AEFe$_2$As$_2$ (AE=Ba or Sr) in the magnetic state\cite{Feng122,HYLiuSr,GDLiuBa,PRichard}. We note that in the N phase of the single-layer FeSe film, this band initially does not cross the Fermi level, with its peak position of photoemission spectra (energy distribution curve, EDC) nearly 25 meV below the Fermi level (sequence 1 in Fig. 2c and Fig. 2f). Further annealing appears to make the N2 band come closer to the Fermi level and eventually get mixed with the electron-like band near M.  (2). The spectral weight distribution as a function of momentum for the N phase shows ``strong spots"-like feature. This is similar to that observed in AEFe$_2$As$_2$ (AE=Ba or Sr) in the magnetic state\cite{HYLiuSr,GDLiuBa}.   However, when coming to the band structure near $\Gamma$, the N phase of the single-layer FeSe is quite different from that of AEFe$_2$As$_2$ (AE=Ba or Sr) compounds. While there are clear hole-like Fermi surface sheets around $\Gamma$ in magnetic BaFe$_2$As$_2$ and SrFe$_2$As$_2$\cite{HYLiuSr,GDLiuBa}, the bands around $\Gamma$ in the N phase are below the Fermi level without Fermi crossing. At present, the exact crystal structure and the physical properties of the N phase remain unclear; we will leave this for the near future investigation.

The single-layer FeSe film on the SrTiO$_3$ substrate exhibits one peculiar electronic characteristic in that the bands near $\Gamma$ point are all pushed below the Fermi level. This is the case for both the S phase and the N phase when they are compared with the hole-like bands near $\Gamma$ in the bulk Fe(Se,Te)\cite{Baumberger} and BaFe$_2$As$_2$ in the magnetic state\cite{GDLiuBa}, respectively. On the other hand, near M point, the electronic structure of the N phase show resemblance to that of the bulk Fe(Se,Te) and the S phase show resemblance to that of the magnetic state of BaFe$_2$As$_2$. Such a disparity in the electronic behavior between the $\Gamma$ and M points in the single-layer FeSe film cannot be explained simply by a rigid band shift due to electron doping in the FeSe film;  it does not happen in bulk Fe(Se,Te)\cite{Baumberger} or Ba(Fe,Co)$_2$As$_2$\cite{Kaminski} systems with similar doping. It points to a mechanism that should be closely related to the SrTiO$_3$ substrate, be it a strain exerted on the FeSe film from the SrTiO$_3$ substrate, or electric polarizability of the dielectric SrTiO$_3$, or the effect of the FeSe$-$SrTiO$_3$ interface. Understanding the origin of this peculiar electronic characteristic of the single-layer FeSe film on SrTiO$_3$ substrate will be important to understand its unusual physical properties.

Now we come to the observation and optimization of the high temperature superconductivity in the S phase.  The tunability of the carrier concentration of the S phase by a simple annealing procedure offers an opportunity to investigate the evolution of its electronic structure and particularly its superconductivity with doping.  We measured superconducting gap and its temperature dependence in a single-layer FeSe film annealed at different sequences (sequence 10 and later) with only the pure S phase (Fig. 3). In this case, the relatively high T$_c$ makes it feasible for us to do the measurements, and it can also avoid complications from another N phase.  To visually inspect possible gap opening and remove the effect of Fermi distribution function near the Fermi level, Fig. 3(a-d) show symmetrized photoemission spectra (EDCs) on the Fermi surface measured at different temperatures, following the procedure commonly used in the study of high temperature cuprate superconductors\cite{MNorman}. The gap opening is characterized by a spectral dip at the Fermi level in the symmetrized EDCs and the gap size is determined by the EDC peak position relative to the Fermi level.   The measured gap size and its temperature dependence, extracted from Fig. 3(a-d),  are shown in Fig. 3(e-h), respectively.  The gap size increases with annealing sequence, from $\sim$10 meV for the sequence 10 to $\sim$19 meV for the sequence 15.  In the mean time, the gap closing temperature  also increases from nearly 40 K for the sequence 10 to nearly 65 K for the sequence 15.  For a given sequence, the dependence of the gap size as a function of temperature basically follows a standard BCS form, as seen from the green lines in Fig. 3(e-h).

The energy gap we have observed in the S phase has been attributed to a superconducting gap.  It is natural to ask whether it can be other kinds of gaps. To pin down the nature of the gap, we have investigated its momentum dependence\cite{DFLiu} and temperature dependence in detail (Fig. 3(i-k) and see Supplementary Fig. S6). We have determined the Fermi momentum both above and below the transition temperature by various methods, including the momentum distribution curves (Fig. 3k), minimum energy gap locus at low temperature, as well as EDC method to determine the dispersion (Supplementary Fig. S6). We find that the Fermi momentum determined by various methods is consistent, and it remains at the same location both above and below the transition. So far, the superconducting gap is the only known case that it opens along the entire Fermi surface and it is symmetric with respect to the Fermi level due to the formation of the Bogoliubov quasi-particles. This is distinct from other transitions like spin-density wave or charge density wave that only the portion of Fermi surface that satisfy the perfect nesting conditions will open a gap on the Fermi surface\cite{GGrunerSDW,GGrunerCDW}; for the rest of the Fermi surface, the Fermi momentum usually exhibits a dramatic difference across the transition temperature\cite{ZXShenBi2201}. Therefore, our observations strongly indicate that the gap we have observed corresponds most likely to a superconducting gap.  Furthermore, the nearly BCS-like form on the temperature dependence of the gap indicates that it is unlikely a pseudogap, as observed in cuprates\cite{Timusk}.  At present, the direct resistivity and magnetic susceptibility measurements on the single-layer FeSe films are still under way due to technical difficulties although the transport measurement on the five-layer FeSe film proved it is superconducting\cite{QKXue}. Our ARPES measurements on the energy gap provide a more direct evidence of the high temperature superconductivity realized in the single-layer FeSe system.

An intriguing issue arises is whether one should treat the single-layer FeSe film on the SrTiO$_3$ substrate as dominantly an isolated 2-dimensional (2D) FeSe system or as a complex interface system with an important interaction between the single-layer FeSe film and the SrTiO$_3$ substrate. In a strict 2D system, one would expect a Kosterlitz$-$Thouless (KT) transition\cite{KTTransition} instead of a true superconducting transition.  Although this question apparently needs further investigations, the fact that the single-layer FeSe film on the SrTiO$_3$ substrate exhibits dramatically different behaviors from the bulk FeSe\cite{MKWu11} and even the FeSe thin films on the graphene substrate\cite{CLSongPRB} indicates that the SrTiO$_3$ substrate and the interface must play an important role.

High temperature superconductivity with a T$_c$ as high as (65$\pm$5) K can be realized in the single-layer FeSe under an optimized annealing condition. As shown in Fig. 3, both the T$_c$$\sim$60 K for the sequence 13 and T$_c$$\sim$65 K for the sequence 15 have exceeded the transition temperature record ($\sim$55 K)\cite{ZARenSm} that has been kept so far in the Fe-based superconductors.  The maximum gap size we have achieved from ARPES measurements ($\sim$19 meV) comes quite close to that measured from the STM$\slash$STS measurements ($\sim$20 meV)\cite{QKXue}.  We note that, although the sample quality gets better with annealing at the initial stage, as signalled by the sharpening of the superconducting quasi-particle peak at low temperature (Fig. 3b and c), further annealing nevertheless makes the quality of the sample deteriorate, as seen from the peak broadening and particularly the signal weakening (Fig. 3d).  This indicates the single-layer FeSe film becomes unstable in the final stage of annealing, which made us unable to go further to achieve even higher dopings.

Fig. 4 schematically summarizes a phase diagram to illustrate the evolution of the electronic structure for the single-layer FeSe film during the annealing process. In particular, the evolution of the superconducting gap ($\Delta$) and the superconducting transition temperature are shown in the pure S phase. Both the superconducting gap and the transition temperature increase with the annealing process, giving an ratio 2$\Delta$$\slash$k$_B$T$_c$ on the order of 6$\sim$7. This indicates that the superconductivity in the single-layer FeSe films is in the strong-coupling regime.

The present work demonstrates that, the electronic structure and physical properties of the single-layer FeSe film can be tuned continuously over a wide range. By a simple mild annealing process, it can even vary from one end phase to the other with different electronic structures. A record high T$_c$ of (65$\pm$5) K can be realized in the system under an optimized annealing condition. Although the present work has not spanned the doping level of the S phase all the way to the over-doped region, the phase diagram obtained looks similar to that of other Fe-based superconductors, and already contains important information for understanding the physics and superconductivity in the Fe-based superconductors. The existence of two distinct phases, plus the tunability of the superconducting properties of the S phase, provide an ideal platform for making heterostructures devices that are needed for both basic studies and potential applications. These include possible combinations of a superconductor with a non-superconductors\cite{IBozovic}, and superconductors with other quantum materials like topological insulators\cite{LFu,JFJiaScience}.

$^{\sharp}$These people contribute equally to the present work.

$^{*}$Corresponding authors: XJZhou@aphy.iphy.ac.cn, qkxue@mail.tsinghua.edu.cn, xcma@aphy.iphy.ac.cn

\vspace{3mm}

\noindent {\bf Acknowledgement} We thank Dunghai Lee and Z.-X. Shen for discussions. XJZ thanks financial support from the NSFC (10734120) and the MOST of China (973 program No: 2011CB921703 and 2011CB605903). QKX and XCM thank support from the MOST of China (program No. 2009CB929400 and  No. 2012CB921702).

\vspace{3mm}

\newpage

\begin{figure}[tbp]
%%\begin{figure*}[floatfix]
\begin{center}
\includegraphics[width=1.0\columnwidth,angle=0]{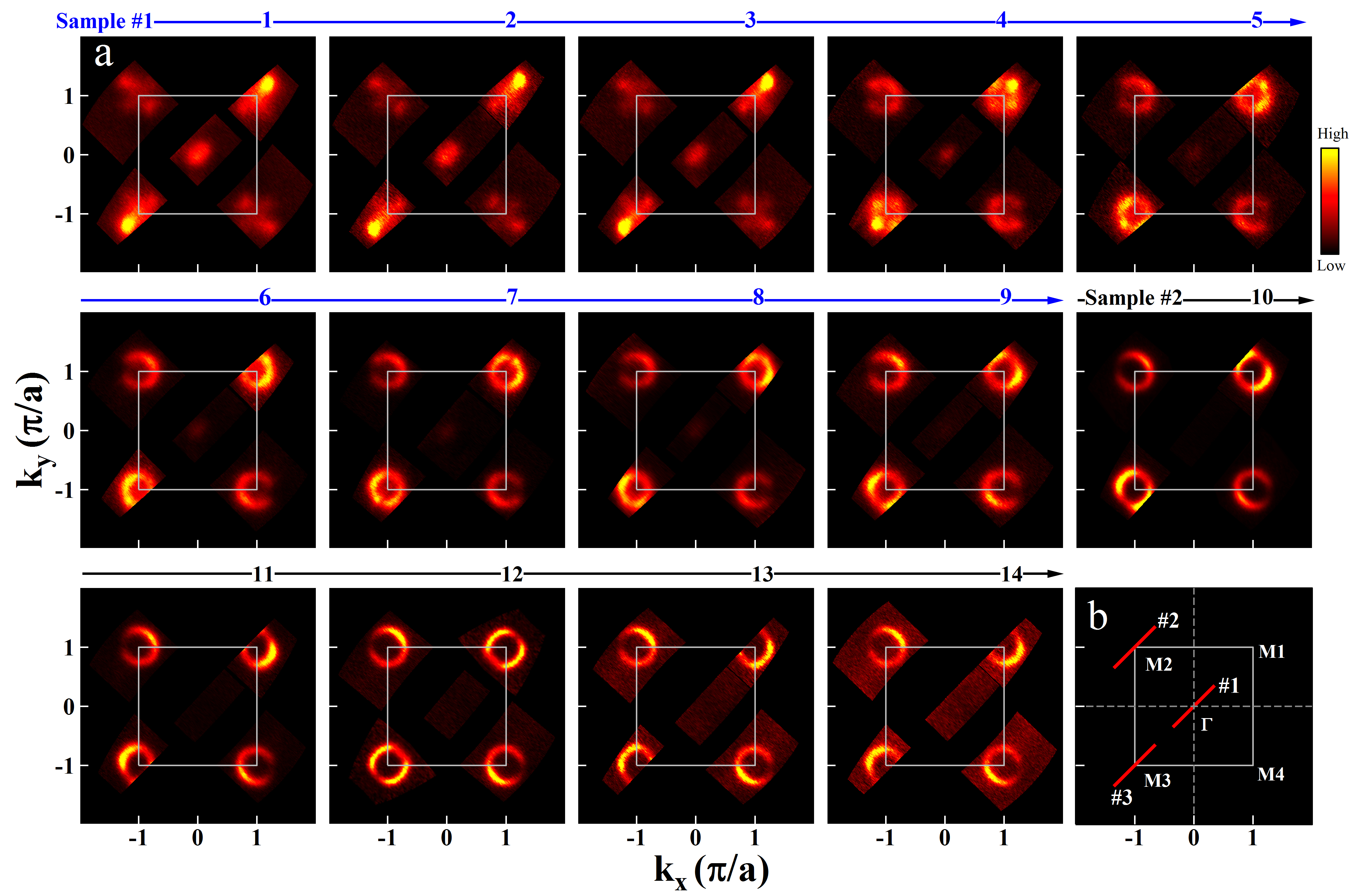}
\end{center}
\caption{Fermi surface evolution of the single-layer FeSe film during the annealing process.  (a). Integrated spectral intensity as a function of momentum for the single-layer FeSe film from two samples annealed at different stages. For convenience, we use the annealing sequence to denote different samples annealed under different conditions, as indicated in the Supplementary Fig. S1. The first 3 images (sequences 1 to 3) are obtained by integrating spectral weight over a small energy window [-0.03eV,-0.01eV] because the observed bands are below the Fermi level.  The rest of images are obtained by integrating over a small energy window [-0.01eV, 0.01eV] around the Fermi level . (b). Labeling of the Brillouin zone; for convenience, the four equivalent M points are labeled as M1($\pi$,$\pi$), M2(-$\pi$,$\pi$), M3(-$\pi$,-$\pi$) and M4($\pi$,-$\pi$). The locations of the momentum cuts for the bands measured in Fig. 2 are also marked.
}
%%\end{figure*}
\end{figure}

\begin{figure*}[tbp]
%%\begin{figure}[b]
\begin{center}
\includegraphics[width=1.0\columnwidth,angle=0]{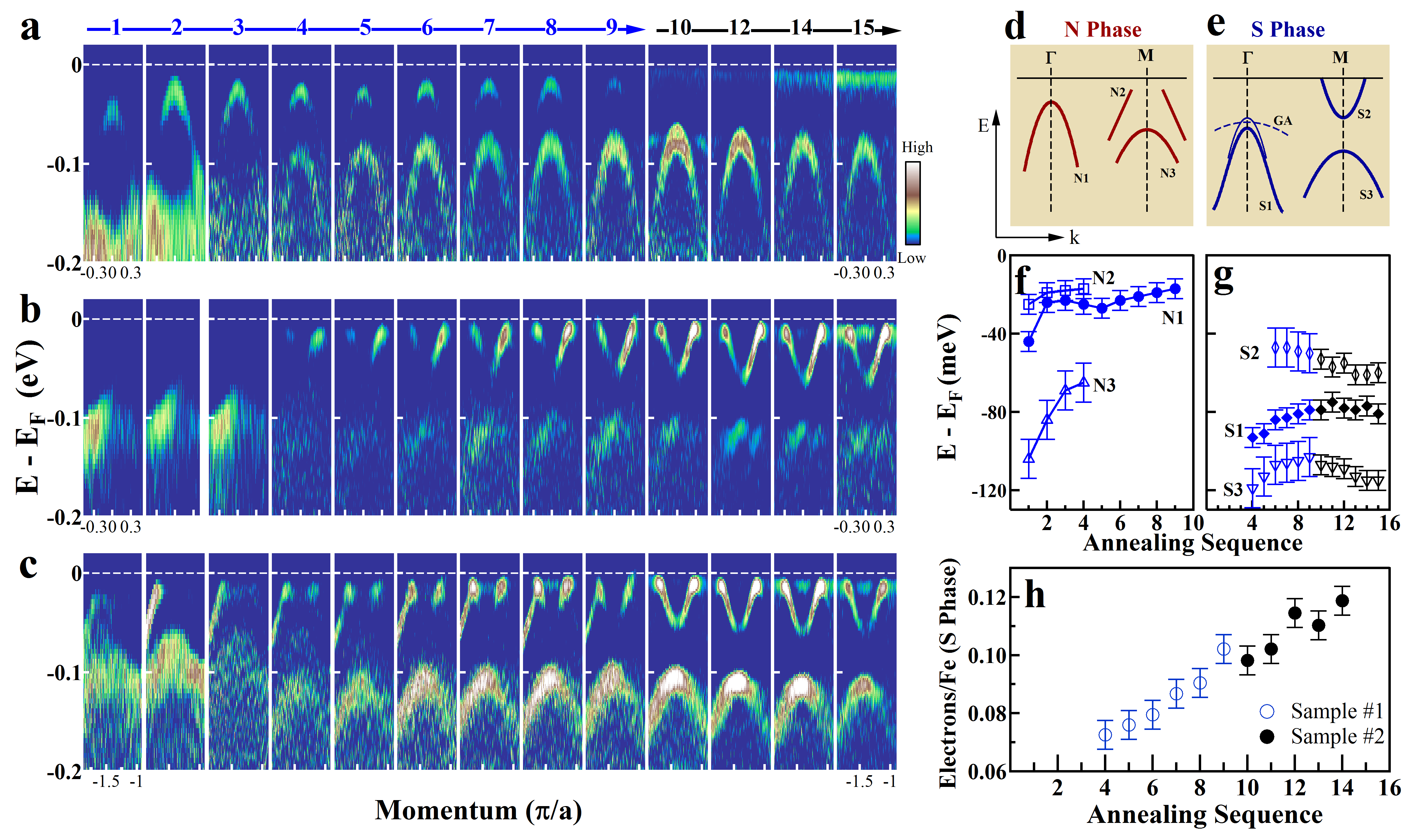}
\end{center}
\caption{Band structure evolution of the single-layer FeSe film during the annealing process. The images are obtained by the second derivative of the original data with respect to the energy. The original bands and associated photoemission spectra for some typical sequences are shown in the Supplementary Figs. S2-S4. (a). Evolution of band structure for the cut $\#$1 near the  zone center $\Gamma$ point in Fig. 1b.  (b). Band structure for the cut $\#$2 near the M2 point in Fig. 1b.  (c). Band structure for the cut $\#$3 near the M3 point in Fig. 1b. (d). Schematic band structure of the N phase. (e). Schematic band structure of the S phase. Near the $\Gamma$ point, in addition to the S1 main band, there is an indication of band splitting near the top of the S1 band, as well as the existence of another GA flat band, as shown in the Supplementary Fig. S3.  (f). Variation of the N1 band top, N2 band tip and N3 band top in the N phase, as shown in Fig. 2d, during the annealing process. (g). Variation of the S1 band top, S2 band bottom and S3 band top in the S phase, as shown in Fig. 2e, during the annealing process. (h). Electron counting for the Fe in the S phase as determined from the electron-like Fermi surface size near M by assuming two degenerate Fermi surface sheets near M.
}
%%\end{figure}
\end{figure*}

\begin{figure}[tbp]
\begin{center}
\includegraphics[width=1.0\columnwidth,angle=0]{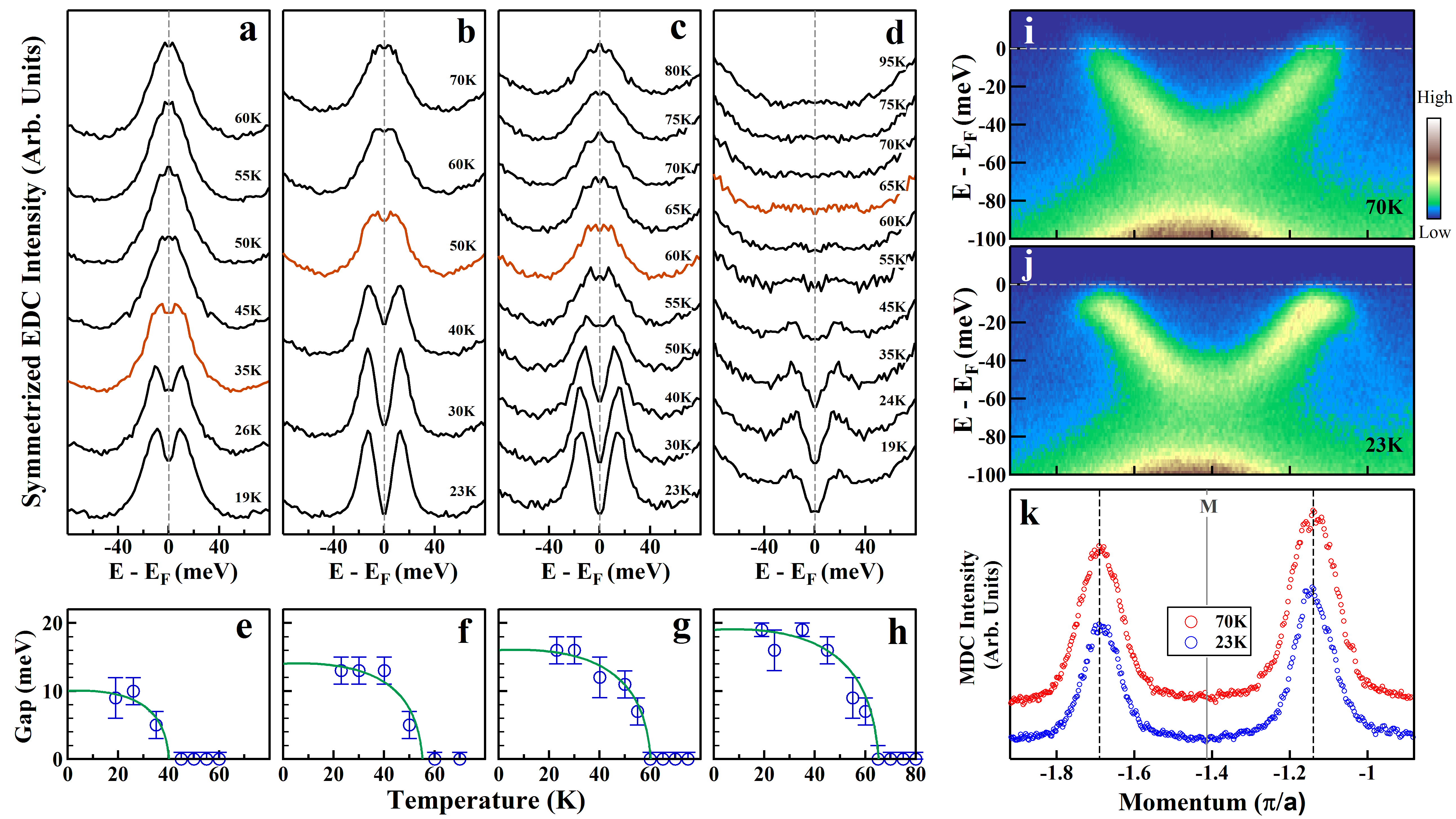}
\end{center}
\caption{Temperature dependence of the energy gap of the single-layer FeSe film annealed under different conditions. (a-d). Symmetrized EDCs on the electron-like Fermi surface near M point at different temperatures for the annealing sequences 10, 12, 13 and 15, respectively. As noted in the Supplementary, the sample temperature has been calibrated. (e-h) shows the temperature dependence of the measured energy gap for the above four sequences. The gap is obtained by picking the position of the symmetrized EDCs relative to the Fermi level. The green lines represent the BCS gap form with different gap size and T$_c$ that can match the measured data. (i-j). Band structure along the cut $\#$3 in Fig. 1b for the sequence 12 measured at 70 K and 23 K, respectively.  The corresponding momentum distribution curves at the Fermi energy for the two measurements are shown in (k). The two MDC peaks show little change in their positions above and below the gap opening temperature ($\sim$ 55 K).
}
\end{figure}

\begin{figure}[tbp]
\begin{center}
\includegraphics[width=1.0\columnwidth,angle=0]{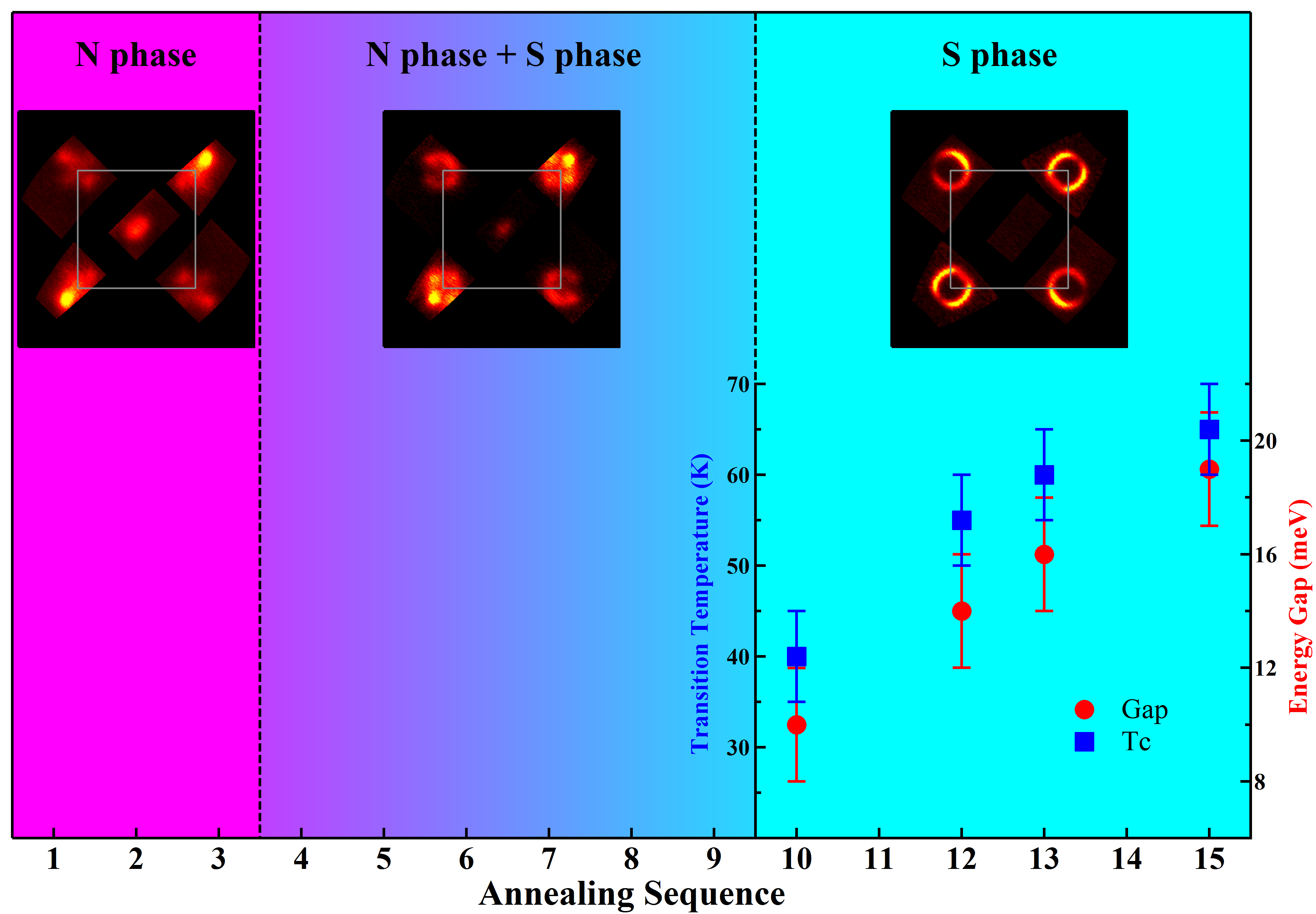}
\end{center}
\caption{Schematic phase diagram of the single-layer FeSe film during the annealing process. Two end phases are identified, with the nearly pure N phase at the initial stage (sequences 1 to 3),  the S phase at the final annealing stage (sequence 10 to 15), and combination of the N phase and the S phase in between (sequence 4 to 9). Representative Fermi surface mappings are shown in the top for the three different regions. In the pure S phase, the measured energy gap size ($\Delta$, solid red circles) and the transition temperatures (T$_c$, solid blue squares) are shown for different sequences; the 2$\Delta$$\slash$k$_B$T$_c$ is near 6$\sim$7.
}
\end{figure}

\end{document}